\documentclass[prb,twocolumn,notitlepage,preprintnumbers,amsmath,amssymb,superscriptaddress]{revtex4-2}
\usepackage{graphicx,bm,amsmath}
\usepackage{amsmath,amssymb}
\usepackage{graphicx}
\usepackage{wasysym}
\usepackage{amsfonts}
\usepackage{bm}
\usepackage{enumerate}
\usepackage{color}
\usepackage{xcolor}
\usepackage{epstopdf}
\usepackage{latexsym}
\usepackage[breaklinks,colorlinks = true,linkcolor = red,urlcolor=cyan,citecolor=red]{hyperref}
\usepackage{stmaryrd} 
\usepackage{braket}

\newcommand{\br}{{\bf r}}

\DeclareMathOperator{\Tr}{\mathrm{Tr}}
\DeclareMathOperator{\rIm}{\mathrm{Im}}

\begin{document}

\title{Direct topological insulator transitions in three dimensions are destabilized by non-perturbative effects of disorder}
\author{Yixing Fu}
\affiliation{Department of Physics and Astronomy, Center for Materials Theory, Rutgers University, Piscataway, NJ 08854 USA}
\author{Justin H. Wilson}
\affiliation{Department of Physics and Astronomy, Center for Materials Theory, Rutgers University, Piscataway, NJ 08854 USA}
\affiliation{Department of Physics and Astronomy, Louisiana State University, Baton Rouge, LA 70803, USA}
\affiliation{Center for Computation and Technology, Louisiana State University, Baton Rouge, LA 70803, USA}
\author{David A. Huse}
\affiliation{Physics Department, Princeton University, Princeton, New Jersey 08544, USA}
\author{J. H. Pixley}
\affiliation{Department of Physics and Astronomy, Center for Materials Theory, Rutgers University, Piscataway, NJ 08854 USA}
\affiliation{Center for Computational Quantum Physics, Flatiron Institute, 162 5th Avenue, New York, NY 10010} 
\date{\today}

\begin{abstract}
We reconsider the phase diagram of a three-dimensional $\mathbb{Z}_2$ topological insulator in the presence of short-ranged potential disorder with the insight that non-perturbative rare states destabilize the noninteracting Dirac semimetal critical point separating different topological phases. Based on our numerical data on the density of states, conductivity, and wavefunctions, we argue that the putative Dirac semimetal line is destabilized into a diffusive metal phase of finite extent due to non-perturbative effects of rare regions. We discuss the implications of these results for past and current experiments on doped topological insulators.
\end{abstract}

\maketitle

\section{Introduction}
The inclusion of topology in understanding the nature of electronic band structures has revolutionized our perspective of materials~\cite{Schnyder-2008,Kitaev2009,Ryu-2010,bernevig2013topological}.  
Following the discovery of three-dimensional (3D) topological insulators~\cite{Liang-2007,Moore-2010} (TIs) in the weakly correlated semiconductors Bi$_{1-x}$Sb$_x$, Bi$_2$Se$_3$,Bi$_2$Te$_3$,and Sb$_2$Te$_3$ (reviews are in Refs.~\cite{Hasan2010,Hasan2011}), a central question has been the stability of these phases to disorder. 
At this point, it is now rigorously established that the 3D $\mathbb{Z}_2$ TI is stable in the presence of disorder~\cite{prodan2016topological} that preserves the protecting time reversal symmetry. 
While Anderson localized Lifshitz states fill the electronic band gap, producing a finite but exponentially small density of states~\cite{RevModPhys.64.755}, the transport gap remains non-zero resulting in a vanishing conductivity, thus converting the system into a topological Anderson insulator~\cite{Sbierski-2014,Kobayashi-2013}. 

\begin{figure}[t!]
\centering
\includegraphics[width=\columnwidth]{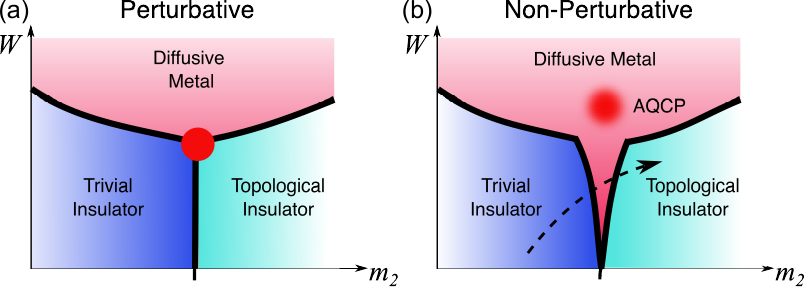}
\caption{
Scenarios for the 
phase diagram of 3D disordered TIs as a function of the Dirac mass parameter $m_2$ and disorder strength $W$. (a) The perturbative scenario, the Dirac semimetal line is stable (though it need not remain a straight vertical line as depicted) to weak disorder and remains a line of critical points separating the two insulating phases. (b) The non-perturbative case, rare regions of the random potential that are ignored in perturbative approaches destabilize the semimetallic critical line into a diffusive metal phase. As a result, the avoided quantum critical point (AQCP) is rounded out and occurs in the diffusive metal phase at a non-universal location that depends on the strength of the avoidance. The dashed arrow denotes a trajectory that different doped samples will take through the phase diagram.
}
\label{fig:shematic_pd}
\end{figure}

In the absence of disorder, it is possible to tune an effective ``Dirac mass'' to induce band inversion within the band structure~\cite{Murakami-2007, Teo2008}.
This realizes a Dirac semimetal critical point (with an odd number of Dirac cones) separating  trivial (or weak TI) and strong TI phases~\cite{Fu-2007,Liu-2010,Imura-2012} as depicted along the Dirac mass $m_2$-axis in Fig.~\ref{fig:shematic_pd}.
There are now a number of experiments that have attempted to tune this mass parameter in various strong spin-orbit coupled insulators by doping the system~\cite{Xu2011,Sato2011,Brahlek-2012,Wu-2013}.
While this does renormalize the mass of the band structure it also introduces disorder into the system.
This raises the question of the stability of the Dirac semimetal critical point in the presence of small but non-zero disorder and the generic structure of the phase diagram of disordered 3D TIs, e.g., does the \emph{clean} critical point remain and evolve into a line as depicted in Fig.~\ref{fig:shematic_pd}(a)?

In the clean limit, the topological critical point realizes a gapless Dirac semimetal, making
the question of the effects of short-ranged quenched disorder surprisingly subtle; the semimetal is more susceptible to the effects of disorder compared to its gapped counterparts~\cite{syzranov2018high,Pixley2021}. 
Disorder is perturbatively irrelevant in a 3D Dirac semimetal~\cite{Fradkin1986,Fradkin1986a,Goswami-2011,Kobayashi2014,Brouwer-2014,Leo-2015,Sergey-2015,Leo-2015,Sbierski2015,Pixley2015,Altland2015,Altland2016,Pixley2016,Bitan-2016,Roy2016,Bera2016,Syzranov2016,Louvet-2016,Sbierski2016,Louvet2017,Luo2018,Luo2018,Balog2018,Roy2018,Brillaux2019,Sbierski2019,Sbierski2020,Kobayashi2020},
so it was originally thought that the semimetallic phase was stable.
However, 
non-perturbative effects arising from rare regions of the random potential destabilize the 3D Dirac semimetal~\cite{Nandkishore2014}. 
Instead it becomes a diffusive metal for any non-zero disorder strength due to rare regions of the random potential creating quasilocalized resonances; 
these rare regions induce a finite density of states at the Dirac node with random matrix theory level-statistics~\cite{Pixley-2016,Pixley-2016a,Pixley2017,Wilson-2017,Holder-2017,Wilson-2018,Wilson-2020,PiresPR-2021,Pires-2022,PiresB-2022}. 
This immediately raises a rather general question as to whether or not the transition between 3D topological and trivial phases is direct (i.e., they are separated by a Dirac semimetallic critical point) or if there is an intervening diffusive metal phase that separates them, as depicted in Fig.~\ref{fig:shematic_pd}(b). The former scenario in Fig.~\ref{fig:shematic_pd}(a) we dub the ``perturbative phase diagram'' as it can be described using the self-consistent Born approximation~\cite{Sbierski-2014,Kobayashi2014}, which is perturbative in the disorder strength. Whereas, the latter scenario [in Fig.~\ref{fig:shematic_pd}(b)] we dub the ``non-perturbative phase diagram'' as it is dominated by the effects of rare regions of the random potential.
The lack of stability of the Dirac semimetal phase~\cite{Pixley2021} implies there is no sharp distinction between the weakly disordered Dirac semimetal and the diffusive metal phase that is produced after the insulating phases have been destroyed, so it is natural to expect that the diffusive metal phase will penetrate all the way down to infinitesimal disorder. 
Demonstrating this diffusive phase with a concrete calculation on a model Hamiltonian for a 3D TI, i.e., distinguishing between Figs.~\ref{fig:shematic_pd}(a) and (b), is the main focus of this work. 

This motivates us to reconsider the effects of disorder on the phase diagram of 3D $\mathbb{Z}_2$ topological insulators.
In order to ascertain the effects of rare regions on the 3D TI phase diagram, we compute the density of states and the DC conductivity on large system sizes (up to a volume of $L^3=200^3$ lattice sites)  by utilizing the kernel polynomial method (KPM)~\cite{Weisse-2006} and handling all matrix-vector multiplication on graphical processing units (GPUs).
As a result, we are able to demonstrate that the density of states and conductivity at the band center remain finite (albeit exponentially small) along the previously expected perturbative semimetal (PSM) line.
This demonstrates the presence of an intervening diffusive metal (whose level statistics at finite energy are those of the Gaussian symplectic ensemble, see Appendix~\ref{sec:A2}), invalidating the perturbative expectations of a vanishing density of states and conductivity.
To study the transitions out of this diffusive metal phase, we turn to analyzing the multifractal spectrum of  eigenstates~\cite{Rodriguez-Multifractal-2011} at the band center.
Thus, we find the non-perturbative phase diagram in Fig.~\ref{fig:shematic_pd}(b) to be the correct physical picture. The size and shape of the intervening metallic phase is non-universal and depends on the choice of disorder distribution and microscopic model.

The remainder of the manuscript is organized as follows: In Sec.~\ref{sec:model} we describe the model we consider and the methods used to compute its properties.
In Sec.~\ref{sec:AC} we discuss the nature of the avoided transition in the model as it appears in the density of states and the conductivity.
In Sec.~\ref{sec:WFs} we use the nature of the eigenfunctions to estimate the localization transitions separating insulating and diffusive metal phases, and we conclude in Sec.~\ref{sec:conclusion}. In Appendix~\ref{sec:A1} we discuss how we suppress finite size effects and in Appendix~\ref{sec:A2} we present level statistics at finite energy.

\section{Model and Approach}
\label{sec:model}
\begin{figure}[b]
\centering
\includegraphics[width=\columnwidth]{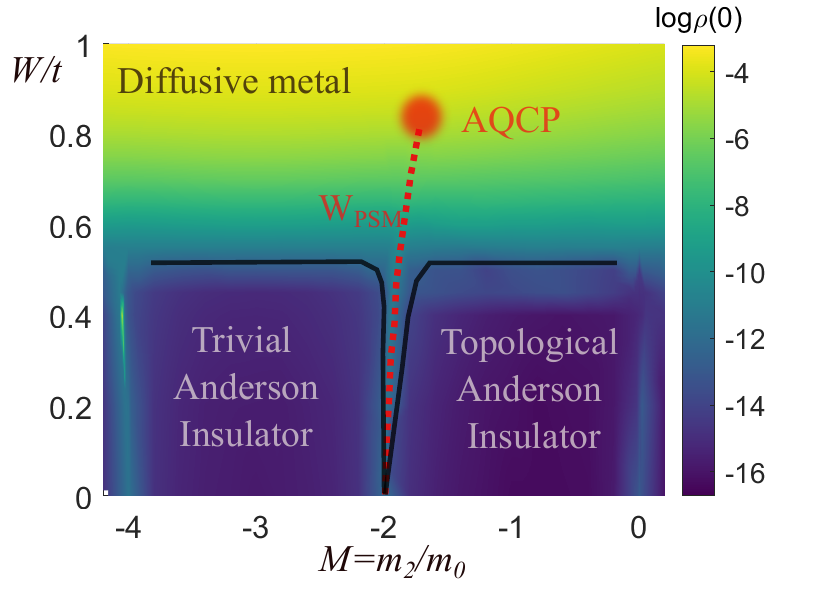}
\caption{Schematic phase diagram labelling the phases, the perturbative semimetal line $W_{\mathrm{PSM}}(m_2)$ (red dashed line), and the avoided quantum critical point (blurred red dot near $W_c(m_2 \approx -1.75)=0.85t$), black lines are drawn for the non-perturbative scenario that is qualitatively consistent with all of the data presented here. 
The color shows $\log \rho(0)$, which is a qualitative proxy for each phase. Density of states $\rho(0)$ is computed with system size $L=151$ with KPM order $N_C=2048$. Each data point is averaged from more than 100 samples with random twisted boundary conditions. The phase boundaries separating Anderson insulating and diffusive metal phases are shown schematically, this will be discussed in detail in Sec.~\ref{sec:WFs}.} 
\label{fig:rho"}
\end{figure}

We study a well known model on a simple cubic lattice in the presence of disorder that realizes a $\mathbb{Z}_2$ TI and has been considered previously~\cite{Kobayashi-2013,Sbierski-2014}, which is defined as
\begin{equation}
   \hat H=\hat H_0+\hat V.
\end{equation}
The topological band structure is due to 
\begin{eqnarray}
    \hat H_0&=&\sum_{\mathbf{r},\mu=x,y,z} \left( \frac{i}{2} t_\mu \psi_\mathbf{r}^\dagger \alpha_\mu \psi_{\mathbf{r} + \hat{\mu}} - \frac{1}{2} m_2 \psi_\mathbf{r}^\dagger\beta\psi_{\mathbf{r}+\hat{\mu}} + {\rm h.c.} \right)
    \nonumber
    \\
    & + & \sum_\mathbf{r} \psi_\mathbf{r}^\dagger \left[ (m_0 + 3m_2) \beta  \right] \psi_\mathbf{r},
    \label{eq:HTI}
\end{eqnarray}
where $t_\mu$ denotes the nearest neighbor hopping strengths  and the topological gap is controlled by the ``mass parameters'' $m_0$ and $m_2$, here we take $t_\mu=t$ to be isotropic for periodic boundary conditions and for twisted boundaries we have $t_\mu \rightarrow t e^{i\theta_\mu/L}$ (where $\theta_\mu \in [0,2\pi)$ is the twist in the $\mu$-direction). In the following we set $t=m_0=1$, and the lattice spacing is also one.
We have  introduced  the four component spinor $\psi_\mathbf{r}$  made of electron annihilation operators $c_{\mathbf{r},\tau,s}$ at site $\mathbf{r}$ with parity $\tau=\pm$, spin $s=\uparrow/\downarrow$, and the Dirac matrices $\alpha_\mu$ and $\beta$ are given by:
\begin{equation}
\alpha_\mu = \tau_x \otimes \sigma_\mu , \quad \beta = \tau_z \otimes \sigma_0.
\end{equation}
For $\hat{V}=0$  this model at the topological transition ($m_2=-2m_0$) has a single Dirac point at 
the $\Gamma$ point (i.e. zero momentum); moving $m_2$ away from this value opens a gap, making a $\mathbb{Z}_2$ topological or trivial (weak in the clean limit) insulator as shown in Fig.~\ref{fig:rho"}.  

The potential disorder is described by 
\begin{equation}
    \hat V=\sum_{\br}\psi_{\br}^{\dag} V(\br)\psi_{\br}
\end{equation}
where $V(\br)$ is a random potential. 
This puts the Hamiltonian in symmetry class AII with $\mathbb Z_2$ topological classification~\cite{Schnyder-2008,Kitaev2009,Ryu-2010,bernevig2013topological}.
Previous studies focused on sampling a box distribution $V(\br) \in [-W,W]$, however bounded distributions are known to suppress the probability to generate rare events~\cite{Pixley-2016a,Nandkishore2014} 
and as a result can artificially mask their effects. Instead, we sample the potential from a Gaussian disorder distribution that has zero mean and standard deviation equal to strength $W$, which is unbounded and known to increase the probability to find rare states.

To solve the model numerically, we compute spectral and transport properties using the KPM. In addition, we study the properties of eigenfunctions near the band center obtained using exact diagonalization or Lanczos. 

To understand the effects on the low energy states near the band center we compute the density of states averaged over disorder samples. This is defined as
\begin{equation}
    \rho(E) = \frac{1}{4L^3}\left[\sum_i \delta(E-E_i) \right]
\end{equation} 
where the linear system size is $L$, the exact eigenvalues $E_i$, and $[\dots ]$ denotes an average over disorder realizations. The DOS is evaluated through expanding this expression in terms of Chebychev polynomials up to an order $N_C$ that is filtered using the Jackson kernel~\cite{Weisse-2006}. The coefficients of the KPM expansion are computed using matrix-vector operations that utilize the recursive nature of the Chebyshev polynomials.

To understand transport properties, we compute the DC conductivity as a function of the Fermi energy at zero temperature using the KPM. Setting $e=1=\hbar$ throughout \footnote{Note that this implies $e^2/h = \frac1{2\pi}$.}, we calculate
the DC conductivity 
using the Kubo formula \cite{garcia2015conductivity}
\begin{equation}
     \sigma(E) = \frac{2}{L^3}\! \int\! f(\epsilon)d\epsilon\rIm \Tr\left [v_x  \frac{dG^-}{d\epsilon} v_x \delta(\epsilon- H)\right ]
     \label{eqn:kubo}
\end{equation}
where $f(\epsilon) = [e^{\beta(\epsilon-E)}+1]^{-1}$ is the Fermi function at inverse temperature $\beta$ and chemical potential $E$ (since we work at zero temperature $E$ is the Fermi energy), $v_x$ is the velocity operator in the $x$-direction,  $G^{-}$ is the retarded Green function, and we average over disorder samples denoted $[\dots ]$.  
We focus on the zero temperature limit and handle the double KPM expansion of Eq.~\eqref{eqn:kubo} that is truncated to an order $N_C$ via GPU's for the matrix-vector multiplication allowing us to reach large system sizes.

To accurately compute the localization phase boundary, we utilize a multifractal finite size scaling approach to wavefunctions from Ref.~\cite{Rodriguez-Multifractal-2011}. We first coarse grain the wavefunction probability density across cubic bins of size $\ell^3<L^3$. After this partition of the system into $(L/l)^3\equiv \lambda^{-3}$ cubes we introduce the probability in cube $k$
\begin{equation}
\mu_k(E) = \sum_{j \in \mathrm{cube} \,\, k}|\psi_j(E)|^2.
\end{equation}
From this a generalized inverse participation ratio $R_q$ and its derivative $S_q=dR_q/dq$ are defined as
\begin{equation}
R_q(E) \equiv \sum_k \mu_k(E)^q, \,\,\,\, S_q(E)=\sum_k \mu_k(E)^q\log \mu_k(E).
\end{equation}
At an Anderson localization transition the wavefunctions become multifractal, this manifests in $R_q$ via the power law dependence on system size
\begin{equation}
\left[ R_q(E) \right] \sim \left( \frac{\ell}{L} \right)^{\tau_q(E)},
\end{equation}
 and the non-linear dependence of $\tau_q$ on $q$ is the hallmark of multifracticality. To compute the location of the localization transitions, we focus on the multifractal spectrum 
\begin{equation}
\alpha_q(E) = \frac{d\tau_q(E)}{dq}.
\label{eqn:alphaq}
\end{equation}
Despite the universal multifractal scaling relations only holding at the critical point, it is useful to extend them to the close vicinity of the transition following Ref.~\cite{Evers-2008}. 
This allows us to estimate $\alpha_q$ in the critical regime via
\begin{equation}
\alpha_q(E) = \frac{[ S_q(E) ] }{ [ R_q(E) ] \log(\ell/L)}.
\label{eqn:alphaq2}
\end{equation}
In the following we use the scaling properties of $\alpha_q$ to estimate the phase boundaries to the Anderson localized phases  between either the trivial or topological Anderson insulating states and the diffusive metal. 
This quantity is particularly useful in the limit of weak disorder that we are focusing on as the spectral gap of each insulating phase will be filled in by an exponentially small contribution coming from non-perturbative and exponentially localized Lifshitz states~\cite{RevModPhys.64.755}. 
It is precisely the delocalization of these states that we are after in the following. 

\section{Avoided criticality}
\label{sec:AC}
As discussed previously, in the clean limit the topological band structure can be tuned through a strong to weak TI transition that is the focus of this work. 
As the weak TI phase has a trivial $\mathbb{Z}_2$ index, 
 we regard this as a phase transition between an Anderson topological insulator and a trivial Anderson insulator  in the presence of finite disorder (though the weak TI phase with disorder remains a rich problem~\cite{Ringel-2012}), see Fig.~\ref{fig:shematic_pd}. 
The critical point separating these two phases in the absence of disorder is a Dirac semimetal with a single Dirac cone in the bulk. 
The lack of stability of the 3D Dirac semimetal to disorder implies that this should become a diffusive metal for infinitesimal disorder,  we explore this regime in the following section.

\subsection{Density of states}

The lack of stability of the Dirac semimetal to disorder is signalled by a non-zero density of states at the Dirac node (which occurs at energy $E=0$).
In the perturbative picture [Fig.~\ref{fig:shematic_pd}(a)], a Dirac semimetal line (shown in Fig.~\ref{fig:rho"}) separates the trivial and topological Anderson insulating phases that terminates at a putative 
tricritical
point separating it from a diffusive metal phase.
In the following, we explore the density of states across the phase diagram and focus in the vicinity of the perturbative Dirac semimetal line that we will show is unstable to disorder due to a finite but exponentially small density of states.
The density of states at the band center across the phase space of $W-m_2$ is a helpful diagnostic for a finite system size~\cite{Wilson-2017} to locate this regime as the density of states is larger here than in the insulating phases. (It should be noted that this diagnostic would work even if the semimetallic phase was stable due to broadening of states in the KPM calculation of $\rho(0)$.)
Thus, we can label the line of maximum density of states at the band center as our estimate of the perturbative Dirac semimetal ``line'';  as a function of $m_2$ it is denoted as $W_{\mathrm{PSM}}(m_2)$, see Fig.~\ref{fig:rho"}.  
$W_{\mathrm{PSM}}(m_2)$ can be computed perturbatively~\cite{Sbierski-2014}, e.g., using the self-consistent Born approximation, however from the non-perturbative perspective this is really more a measure of the center of the disorder-induced diffusive metal phase.
To clearly identify this regime as a diffusive metal we also investigate the DC transport properties and the nature of the wavefunctions in the following.
The level statistics showing the diffusive metal is consistent with the Gaussian Symplectic Ensemble (GSE) random matrix theory ensemble are shown in Appendix~\ref{sec:A2}.

\begin{figure}[t!]
\centering
\includegraphics[width=\columnwidth]{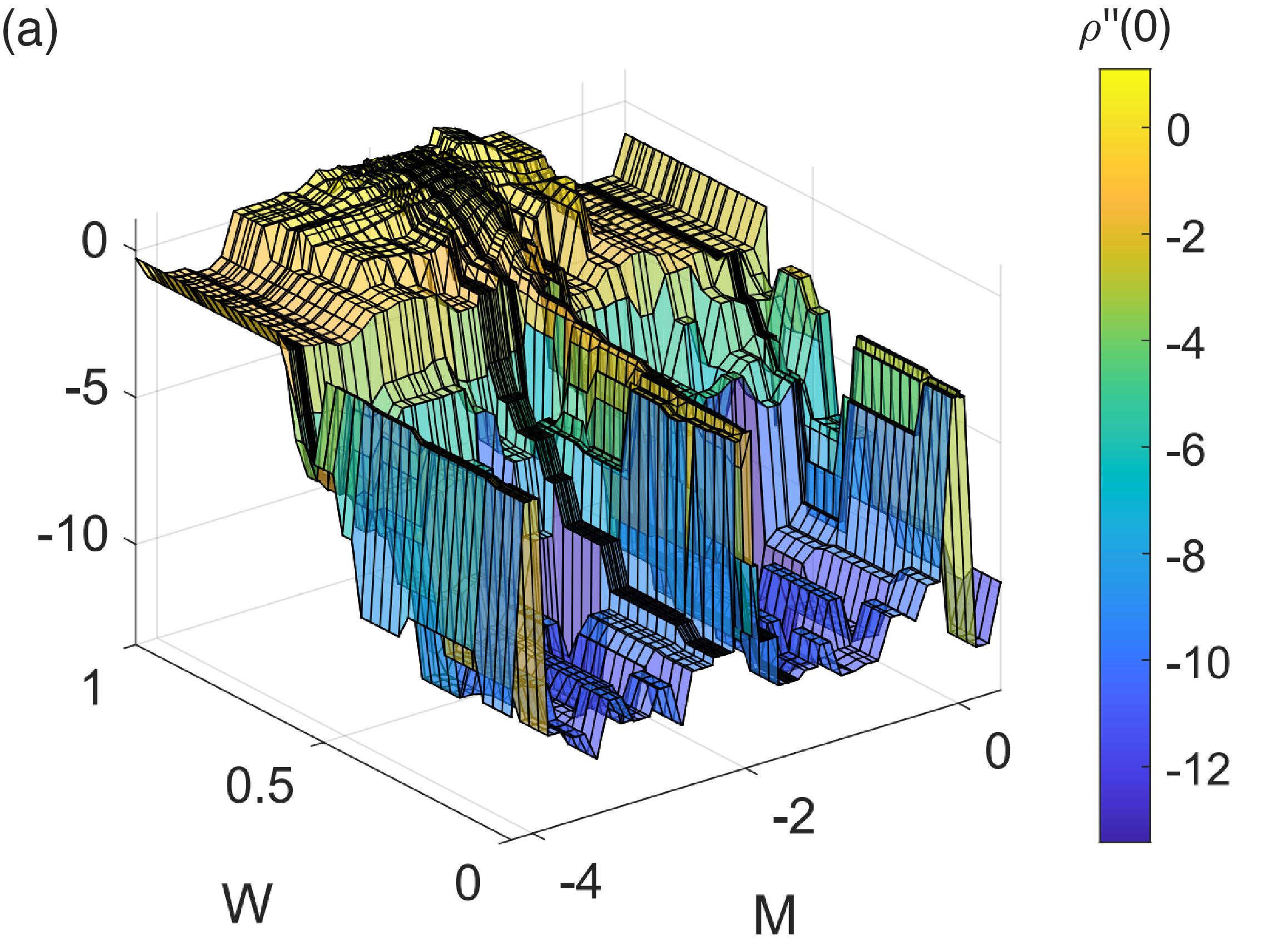}
\\
\includegraphics[width=\columnwidth]{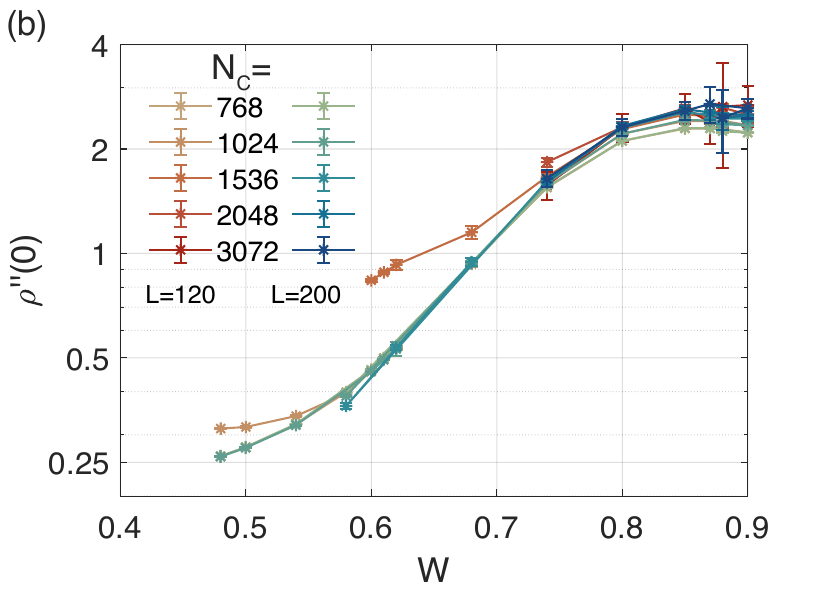}
\caption{{\bf Avoided transition seen through a converged $\rho''(0)$}: (a) A map of the second derivative of the density of states across the phase diagram in $W$ and $M=m_2/m_0$ is shown for system size $L=200$ and KPM expansion order $N_C=2048$. 
The appearance of the PSM is clear from the ``ridge'' of local maxima in $\rho''(0)$ near $M=-2$. 
(b)  We show the average second derivative of the density of states for several KPM expansion orders and two system sizes $L=120$ and $L=200$ 
along the PSM line, showing that the peak is converged in $L$ and $N_C$  and that it has only a weak maximum roughly around $W_c(m_2=-1.75)\approx 0.85$.  
Beyond $W>0.9$  the PSM can no longer clearly be found  but $\rho''(0)$ decreases for larger $W$ indicating the weak maximum as shown in the top figure.
}
\label{fig:DOS_sm}
\end{figure}

The rounding of the perturbative critical point along the putative Dirac semimetal line  can be probed through the analytic properties of the density of states. Assuming the transition is avoided allows us to Taylor expand the energy dependence of the density of states (along this line):
\begin{equation}
    \rho(E) = \rho(0) + \frac{1}{2}\rho''(0)E^2 + \frac{1}{4!}\rho^{(4)}(0)E^4 + \dots 
\end{equation}
and if this assumption is invalid the density of states will become non-analytic signalled by a divergence in the derivatives of the density of states. 
As has been shown in several lattice models of Dirac and Weyl semimetals, the avoided quantum critical point can be located by the maximum in $\rho''(0)$ as a function of $W$, and the strength of avoidance is measured by the size of this peak once it has been saturated in system size and KPM expansion order~\cite{Pixley-2016, Pixley-2016a}.

To establish the location of the avoided transition, denoted $W_{\mathrm{AQCP}}$, we compute $\rho''(0)$ across the phase diagram in the space of $W-m_2$ as shown in Fig.~\ref{fig:DOS_sm}(a). 
We find a broad maximum in the space of $W-m_2$ and estimate the AQCP from where it is maximal along the the line defined by $W_{\mathrm{PSM}}(m_2)$, which yields $W_{\mathrm{AQCP}}(-1.75\pm 0.05)/t=0.85\pm 0.05$ 
though the peak there in $\rho''$ vs.\ $W$ is quite weak.
As we show in Sec.~\ref{sec:ACcond} below, this estimate of the location of the AQCP is consistent with the appearance of a critical scaling regime of the conductivity.
To demonstrate the transition is avoided in the thermodynamic limit, we consider two larger system sizes ($L=120,200$) and saturate the peak of $\rho''(0)$ in  
the KPM expansion order ($N_C$) as shown in Fig.~\ref{fig:DOS_sm}(b), 
demonstrating the density of states remains a smooth function in this regime, and the putative transition is avoided.

\begin{figure}[t!]
\centering
\includegraphics[width=\columnwidth]{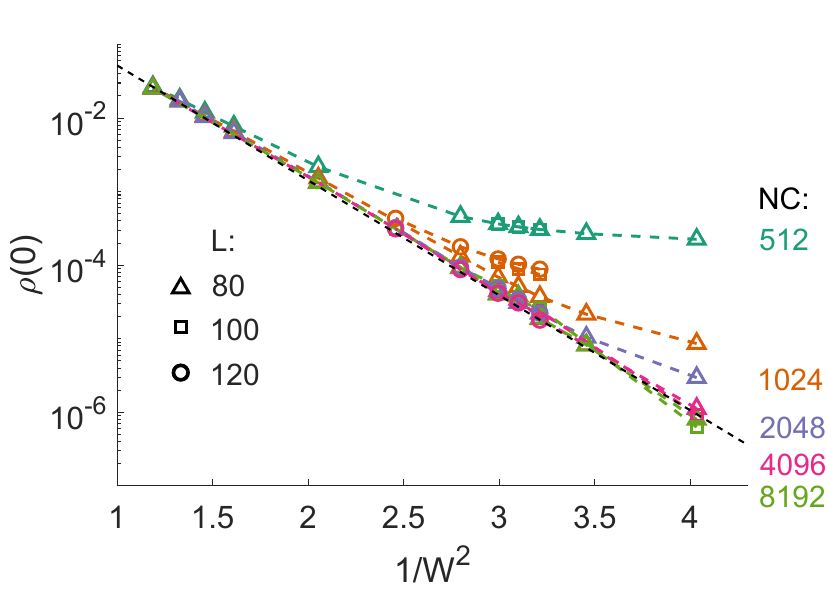}
\caption{{\bf Density of states along the perturbative semimetal line}: We show the average density of states at $E=0$ for three systems sizes (denoted by the symbol type) and several KPM expansion orders along the PSM line, showing that it is converged in $L$ and $N_C$ at these values of $W$ and that it follows the non-perturbative rare region form $\log \rho(0) \sim -(t/W)^2$. 
}
\label{fig:DOS_sm_line}
\end{figure}

Having located the avoided transition we can now ``follow'' the density of states down the putative semimetal line. 
As shown in Fig.~\ref{fig:DOS_sm_line}, we find the density of states is non-zero but becomes exponentially small at weak disorder and we find it follows the rare region form 
\begin{equation}
   \log \rho(0) \sim -\left(\frac{t}{W_{\mathrm{PSM}}(m_2)}\right)^2
   \label{eqn:rhoRR}
\end{equation}
along the PSM following $W_{\mathrm{PSM}}(m_2)$. 
This demonstrates that the density of states is non-zero along this line stretching down to weak disorder.
Importantly, we are able to converge $\rho(0)$ in system size and KPM expansion order, and the resulting data follows Eq.~\eqref{eqn:rhoRR} across close to five orders of magnitude in $\rho(0)$.

The presence of nearby Anderson insulating states makes identifying the presence of the diffusive metal phase solely through the density of states insufficient as localized Lifshitz states occur away from the mobility edge and contribute to a finite density of states.
Therefore, we now turn to computing the DC conductivity along this perturbative Dirac semimetal line paying particular attention to where we demonstrated the nonzero density of states.

\subsection{Conductivity}
\label{sec:ACcond}

\begin{figure}[t]
\centering
\includegraphics[width=\columnwidth]{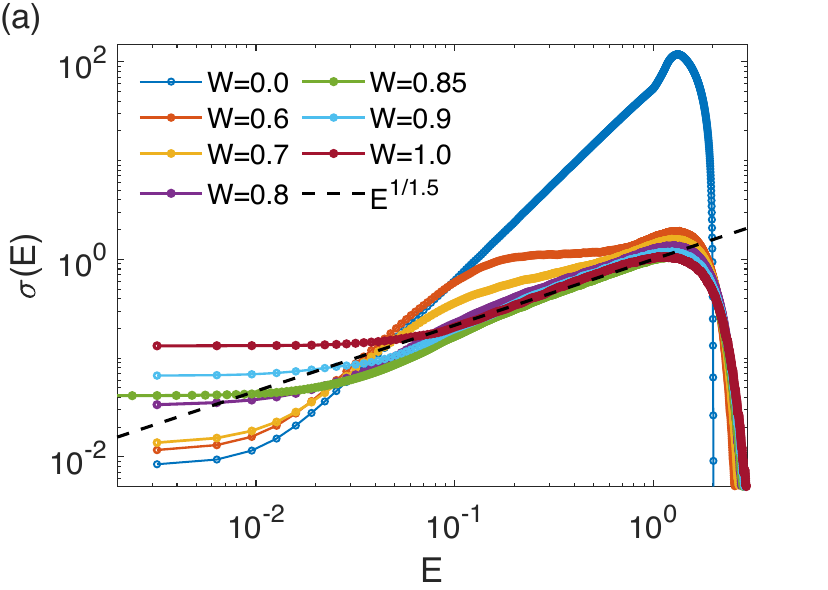}
\includegraphics[width=\columnwidth]{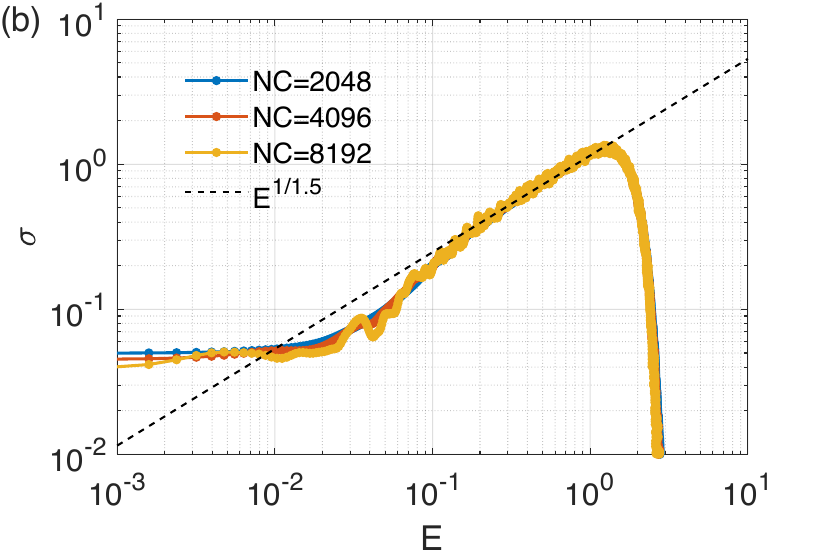}
\caption{{\bf Observing the AQCP in the scaling of the conductivity}: The DC conductivity as a function of the Fermi energy $E$ computed with KPM.  (a) for several values of the disorder strength $W$ and $N_C=2048$ with $L=85$. Near the avoided transition found from the peak $W_{\mathrm{PSM}}(m_2\approx -1.75)\approx 0.85t$ in $\rho''(0)$ we find that above a low energy cross over scale ($E>E^*$) the conductivity scales like $\sigma \sim E^{1/z}$ with $z\approx 1.5$ in excellent agreement with the expectation of $z$ based on the known nature of the AQCP. (b) shows for $W=0.85$ and $M=-1.8$ at various larger $N_C$ that by increasing $N_C$, the scaling does not change and the low energy roll off at the crossover scale $E^*$ is due to a nonzero DC conductivity in the zero energy limit.
}
\label{fig:con_aqcp}
\end{figure}
We now turn to the DC conductivity  at zero temperature.
Having demonstrated the presence of an AQCP in the density of states we now consider how signatures of the perturbative transition show up in the scaling of the conductivity. If the transition was not avoided then the DC conductivity at the transition should vanish 
at $E=0$ as a power of $|E|$~\cite{Sbierski2015}, but we find this is rounded out for $W\approx W_{\mathrm{AQCP}}(m_2=-1.75t)=0.85t$, in particular a finite energy cross over scale (that is generated by non-perturbative effects) exists ($E^*$), such that at energies above it the conductivity looks critical, namely
\begin{equation}
    \sigma(|E| \gg |E^*|)\sim |E|^{1/z}.
\end{equation}
 We find that  $z\approx 1.5$, as shown in Fig.~\ref{fig:con_aqcp}, in excellent agreement with previous estimates of $z$ at the AQCP~\cite{syzranov2018high,Pixley2021}.
 However, as the transition is avoided and the phase at $E=0$ is a diffusive metal [$\sigma(E=0)>0$], we expect that at low enough energies ($|E|<|E^*|)$ this scaling will be spoiled by the finite value of $\sigma(0)$ as $E\rightarrow 0$ defining a cross over scale $|E^*|>0$.
 In Fig.~\ref{fig:con_aqcp}(b), $E^*$ appears as the rounding of the conductivity from power-law to essentially $E$-independent at small $|E|$.

\begin{figure}[b]
\centering
\includegraphics[width=\columnwidth]{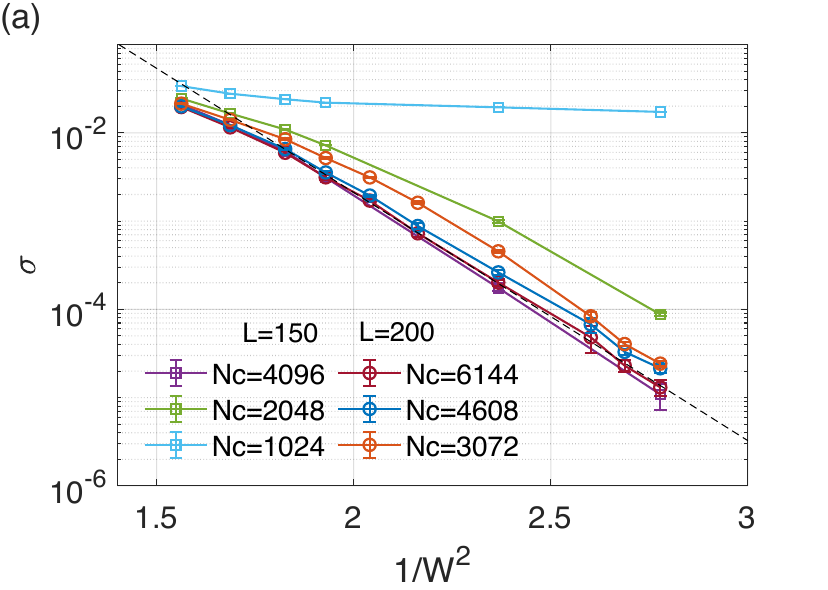}
\includegraphics[width=\columnwidth]{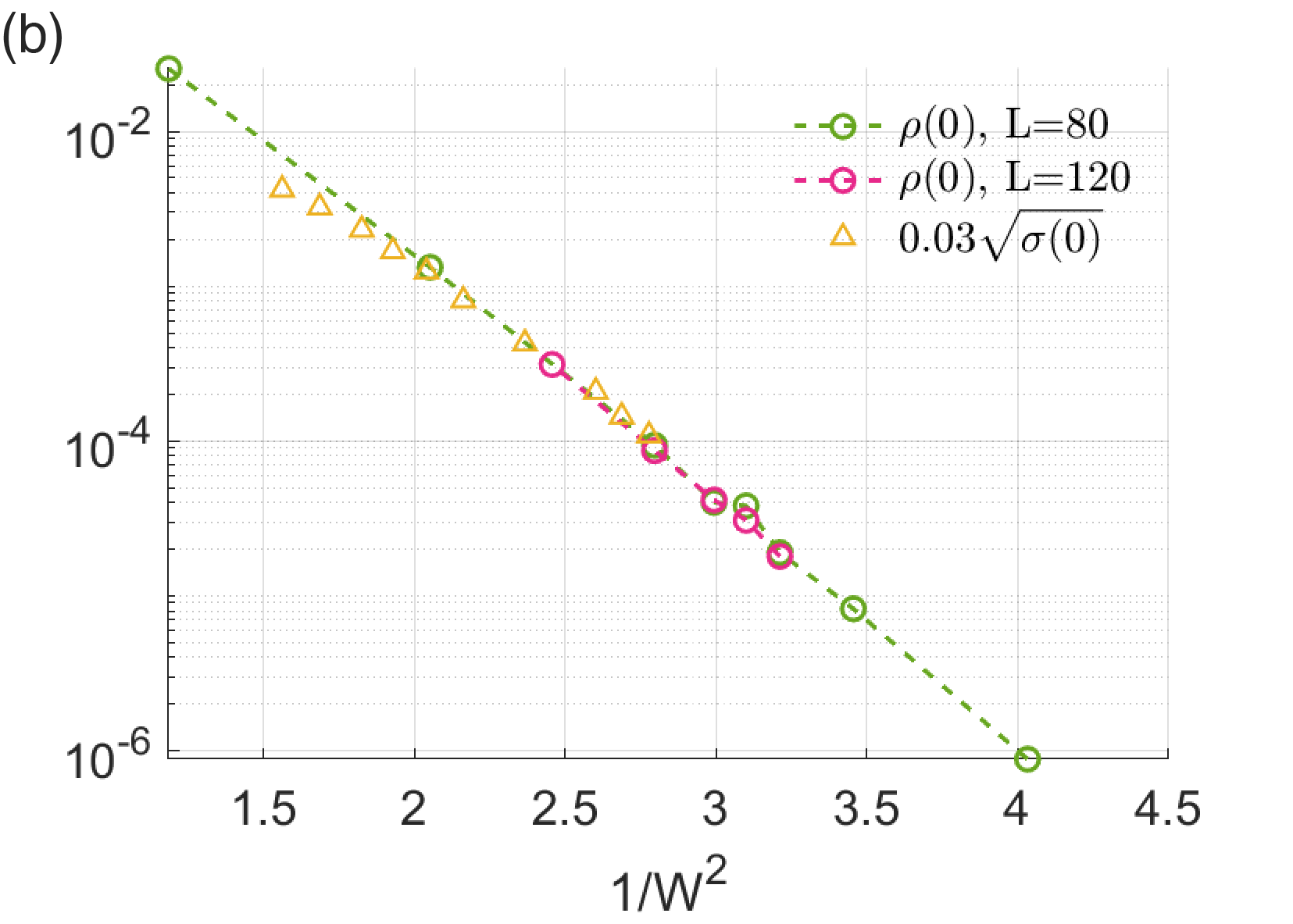}
\caption{ {\bf A diffusive metal with a finite conductivity along the PSM line}. (a) Along the line $W_{\mathrm{PSM}}(m_2)$ we are able to converge the DC conductivity at the largest system sizes ($L$) and ($N_C$) down to an order $\sim 10^{-5}$. 
While it is much more challenging to converge this as far along the line as the density of states in Fig.~\ref{fig:DOS_sm}, we are still able to find a converged DC conductivity down to $W=0.6t$, well below our estimate of the AQCP at $W=0.85t$. (b) Comparison of the scaled conductivity ($\sqrt{\sigma(0)}$) for $L=200$ and $N_C=6144$ and the density of states $(\rho(0))$ for $L=80$ and $L=120$  with $N_C=4096$ in the rare region regime yielding $\rho(0)\approx 0.03 \sqrt{\sigma(0)}$. 
}
\label{fig:con_sm}
\end{figure}

To extract the rare-region contribution to the DC conductivity, we find it useful to use twisted boundary conditions with a twist of $\bm{\theta}=(\pi,\pi,\pi)$ with even system size $L$ to induce the largest possible finite size gap in $\sigma$. 
This allows the rare state contribution to populate the finite-size gap (similar to what has been successful for the density of states~\cite{Pixley2015}).
We average $\sigma$ over 100 samples for $L=150$ and 50 samples for $L=200$, 
utilizing large system sizes and KPM expansion orders enabled by our GPU implementation.
To remove the leading perturbative, finite-size effect, we ensure that each random sample that has a potential that sums exactly to zero by shifting the random potential by its average. 
We discuss these effects in more detail in Appendix~\ref{sec:A1}.

Using this approach, we converge the rare-region contribution to the DC conductivity along the PSM line in system size and KPM expansion order.
Our results are shown in Fig.~\ref{fig:con_sm}(a).
We see that at disorder strengths on the order of the AQCP ($1/W^2\approx 1.3)$ the conductivity is well converged at small $N_C$ and $L$, as $W$ decreases along the perturbative semimetal line, we see the data remains converged at our largest system size $L=200$ and expansion order $N_C=6144$ down to weak disorder strengths, well below the AQCP to $W\approx 0.6t$.
We find in the band center (and in the regime $0\le E<E^*$) that the converged DC conductivity is exponential small and follows the non-perturbative rare region form similar to the density of states
\begin{equation}
  \log  \sigma(E=0) \sim - \left(\frac{t}{W_{\mathrm{PSM}}(m_2)}\right)^2.
  \label{eqn:condRR}
\end{equation}
Thus, our results are consistent with the entire PSM line being a diffusive metal \emph{phase} with a nonzero density of states and DC conductivity due to rare-regions of the random potential. 
Comparing the fits of the rare-region functional forms in Eqs.~\eqref{eqn:rhoRR} and \eqref{eqn:condRR}, namely $\log \rho(0) \approx -a/W^2 + b$ and $\log \sigma(0) \approx -a'/W^2 + b'$ yields $a=3.6$ and $b=0.7$ for the density of states and $a'=6.4$ and $b'=6.7$ for the conductivity, which approximately yields the relationship $\sigma(0) \approx (\rho(0)/0.03)^2$ in the rare region dominated regime as shown in Fig.~\ref{fig:con_sm}(b).

Thus, we conclude that in the rare-region dominated regime, at weak disorder, transport is facilitated by tunneling between these rare regions with large probability amplitude resulting in
\begin{equation}
    \sigma(0)\sim \rho(0)^2.
    \label{eqn:transportRR}
\end{equation}
In this regime, the diffusion constant apparently behaves as
\begin{equation}
    D=\sigma(0)/ \rho(0) \sim \rho(0), 
    \label{eqn:diffusion}
\end{equation} 
which is also exponentially small and follows Eq.~\eqref{eqn:rhoRR}.
These results represent the first direct demonstration that the rare region dominated regime yields diffusive transport properties, which is a central result of this manuscript. It is interesting to compare this result with a self-consistent $T$-matrix calculation~\cite{Holder-2017}, which obtained
an effectively constant diffusivity, in contrast to our finding in Eq.~\eqref{eqn:diffusion}.  We have not developed any theoretical understanding of this difference, and leave this question for future work.

By varying the mass parameters of the model we can 
tune the system out of this diffusive metal phase at weak disorder into either topological or trivial insulating phases, which we now turn to.

\section{Anderson localization transitions}
\label{sec:WFs}
 In the following section, we explore the Anderson localization transitions in close proximity to the diffusive metal regime that we have identified along the perturbative semimetal line.
 Starting in either insulating phase and turning on a weak disorder potential will fill in the spectral gap, but the density of states and the conductivity will remain exponentially small (in the disorder strength $W$),
 which makes a precise estimate of the conductivity a challenging computational task.
 Due to this, we find that the KPM approach to the DC conductivity has trouble precisely locating the localization phase boundaries, as the finite KPM-expansion order broadens the low-energy and finite-size scaling of the DC conductivity.  As a result, to provide a separate identification of the Anderson localization phase boundaries we systematically study the nature of eigenfunctions near the band center across the transition, as described in Eq.~\eqref{eqn:alphaq}. 
 Because the density of states is so small here, eigenfunctions near the band center can be obtained efficiently using Lanczos based approaches and thus computed over a large number of samples that allow us to estimate the Anderson insulator (trivial and topological) to diffusive metal transition over a narrow energy window at the band center.

 \begin{figure}[t]
\centering
\includegraphics[width=\columnwidth]{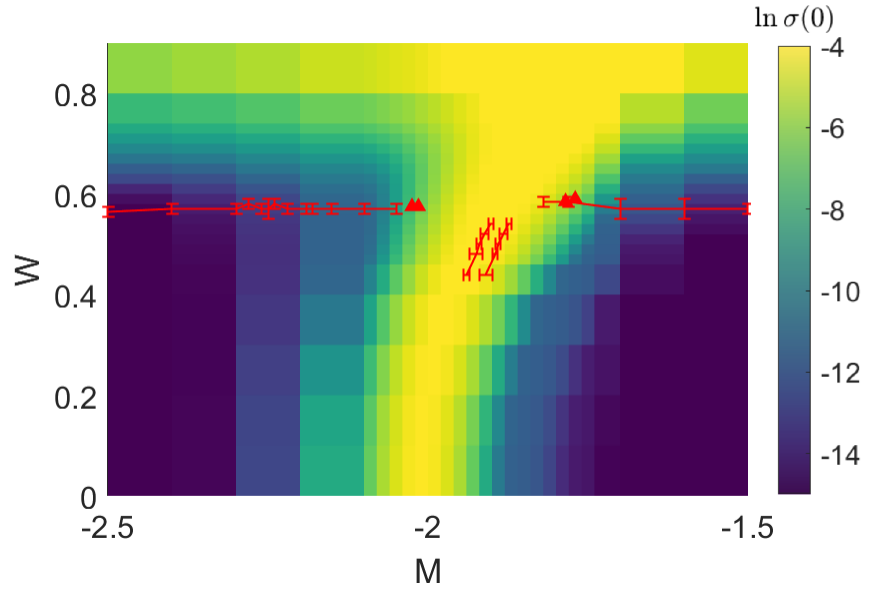}
\caption{{\bf Anderson localization phase boundaries}. 
A zoom into the region near the transition, the color is the value of $\log\sigma(0)$ at the band center for $L=101$ and $N_C=1024$ displaying the insulating phases (topological for $M=m_2/m_0>-2$ and trivial for $M<-2$) in blue and the metallic phase in yellow. 
Based on the multifractal properties of the eigenfunctions near the band center, we use the crossing of the two largest sizes to estimate a bound for the localization transition shown in red, see Fig.~\ref{fig:alpha_q}.  For the vertical cuts along $W$ this acts as an upper bound on the transition and importantly for the data points in the yellow region of the color plot the crossings drift outward away from the center of the metallic wedge. Taken together, we are able to discern the presence of a finite but narrow metallic phase between the two red phase boundaries.  While this becomes even more narrow at weak disorder, our data is consistent with the non-perturbative scenario in Fig.~\ref{fig:shematic_pd}.
}
\label{fig:con_pd}
\end{figure}

 We first show, in Fig.~\ref{fig:con_pd}, the conductivity across the phase diagram on a log-scale that more clearly shows a finite diffusive metal \emph{phase} separating two regimes with a vanishingly small conductivity. 
 We explore several cuts across this phase diagram in this section.

\begin{figure*}[t!]
\centering
\includegraphics[width=0.675\columnwidth]{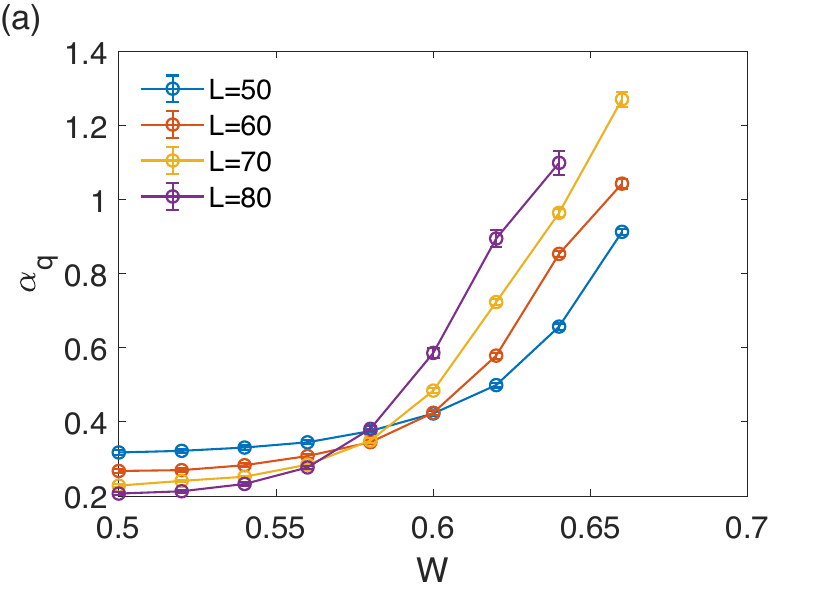}
\includegraphics[width=0.675\columnwidth]{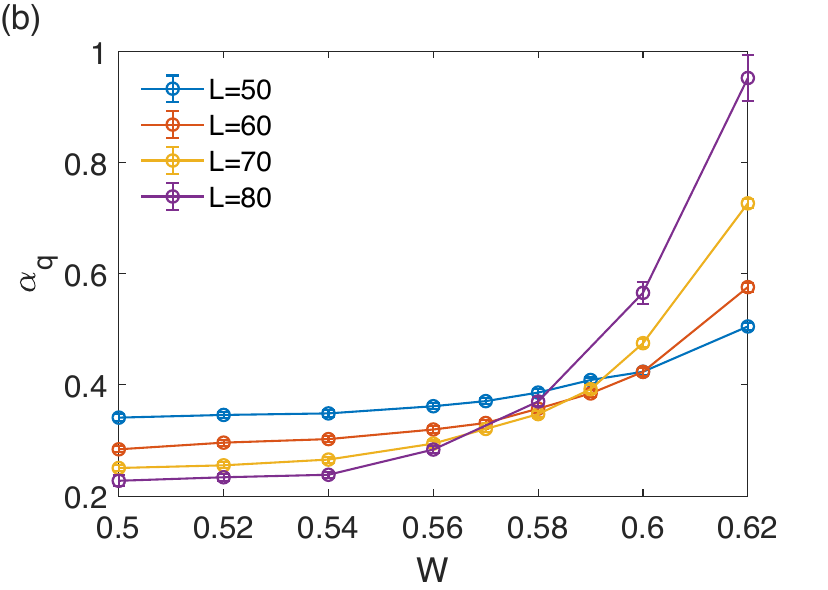}
\includegraphics[width=0.675\columnwidth]{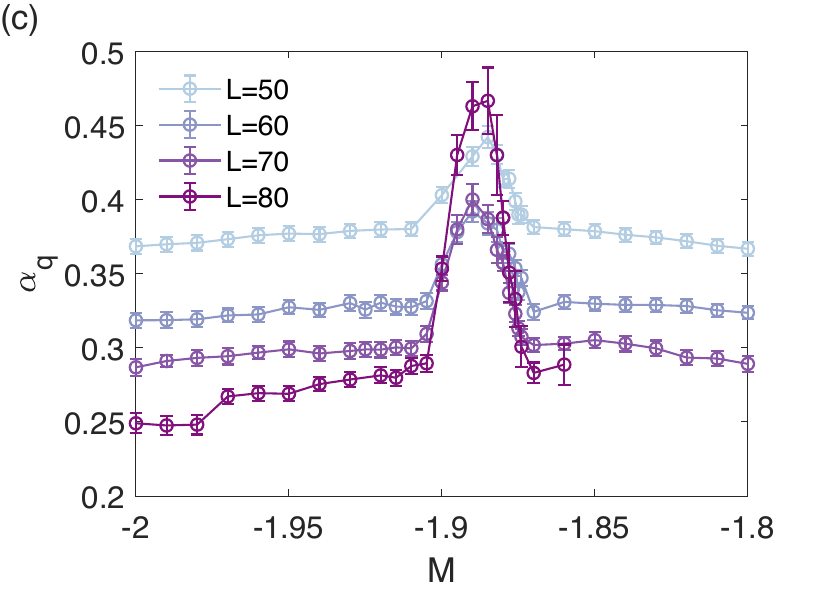}
\caption{{\bf Wave functions near the band center exhibiting the localization transition through multifractral finite-size scaling.}  At fixed $M=m_2/m_0$ we consider a cuts along the topological-Anderson-insulator to diffusive-metal transition (a) at $M=1.6$, and along the trivial Anderson insulator to diffusive metal transition (b) for $M=2.1$. We expect that $\alpha_q$ will cross for several system sizes at the transition. Instead, we see a clear drift in the crossing as we increase the system size. This demonstrates there are large finite size corrections to the crossing at these sizes. 
Therefore, we take the crossing between the largest pair of system sizes as an estimate of the bound (from above) of the localization transition.
(c) A cut at fixed $W=0.54$ as a function of $M$. 
Here we identify two separate crossings in $\alpha_q$ demonstrating the presence of \emph{two separate} transitions. We also find a clear drift in these crossings, here the drift is importantly \emph{outward} away from the perturbative semimetal line. 
}
\label{fig:alpha_q}
\end{figure*}

To estimate the localization transition at the critical disorder strength $W_l$ and energy $E$ near the band center (practically, we take a small but finite energy window to be $|E|<0.001$) as a function of $W$ we use the finite size scaling ansatz on $\alpha_q$ from Eq.~\eqref{eqn:alphaq2} to obtain
\begin{equation}
\alpha_q(E)\sim g_q(E,|W-W_l|L^{1/\nu}),
\end{equation}
where $g_q$ is an unknown scaling function and $\nu$ is the localization length exponent. This ansatz implies the data on $\alpha_q$ for various system sizes will cross at $W_l$ allowing for an unbiased estimate of the localization transition. 
This makes this object more useful then the inverse participation ratio as the latter has an overall scaling dimension which results in a more complicated finite size scaling analysis to estimate $W_l$.

We present two distinct cuts across the phase diagram in Fig.~\ref{fig:rho"} from an Anderson topological insulator to a diffusive metal in Fig.~\ref{fig:alpha_q}(a) and a trivial Anderson insulator to diffusive metal in Fig.~\ref{fig:alpha_q}(b). 
As can be seen from the data, there is a clear drift in the crossing of each pair of increasing system sizes. Due to the large drift and the available system sizes used in the numerics we are unable to provide an accurate estimate of $W_l$. Instead we use the location of the crossing between the two biggest system sizes as an upper bound on $W_l$ as the drift in the crossing is towards smaller $W$.

In Fig.~\ref{fig:alpha_q}(c), we present a cut as a function of $M=m_2/m_0$ at fixed disorder strength well below the AQCP.
We first start at weak disorder strength ($W=0.54t$), where the perturbative picture would predict a \emph{direct transition} between a topological and trivial Anderson insulator. 
However, this is inconsistent with  our results. 
Instead, we find that on this range of system sizes there are two clear crossings, with a clear drift outward away from the PSM line on the two sides of the diffusive metal ``sliver'' in Fig.~\ref{fig:con_pd}, providing strong evidence for the appearance of an intervening delocalized phase. 
We then follow this upwards along the PSM line [e.g. following $W_{\mathrm{PSM}}(m_2)]$  allowing us to track the two separate estimates of the phase boundary bounds that are placed with red data points in the conductivity color map in Fig.~\ref{fig:con_pd}. 
This intervening delocalized phase is precisely the diffusive metal phase penetrating down the space between the two insulating phases that we have previously identified with the converged DC conductivity that is exponentially small in Fig.~\ref{fig:con_sm} and described by Eq.~\eqref{eqn:condRR}. 
This in conjunction with the converged DC conductivity on large system sizes provide numerical evidence for the absence of a direct transition between trivial and topological insulating phases in the presence of disorder.

 \section{Discussion and Conclusion}
In this work, we have explored the possibility of an intervening diffusive metal phase separating 3D trivial and topological Anderson insulators at weak disorder. 
By following the line of maximal density of states on finite size simulations, we are able to track the perturbative semimetal ``line'' that ends with a strongly avoided transition at larger disorder strength.
Along this line, we are able to converge the density of states and DC conductivity to a non-zero but exponentially small value down to disorder strengths well below the estimate of the avoided transition. 
To the best of our knowledge, this work provides the first estimate of the exponentially small but finite rare-region contribution to the DC conductivity below the avoided transition.
To ascertain the location of the Anderson transitions of the trivial and topological insulating phases, we used the finite-size multifractal scaling of the wavefunctions to provide strong evidence of two separate localization transitions due to the diffusive metal phase penetrating down the phase diagram below the avoided transition. 

Our work points to a strong dichotomy between doping tuned topological transitions and those that are tuned by pressure or optical means.
In fact, disorder introduced by doping will lead to an intrinsic broadening of the expected semimetal point into a metallic phase and experimentally it is expected to be represented as a regime of finite extent.
Indeed in experiments on BiTl(S$_{1-\delta}$Se$_\delta$)$_2$, the finite regime of doping $0.4<\delta<0.6$ was identified as separating  the two trivial and topological insulating phases~\cite{Xu2011},  directly in line with the expectations based on our results.
In contrast, pressure tuned or optically activated topological phases~\cite{aryal2022robust} in nominally undoped samples should have a very narrow metallic phase being set by the intrinsic disorder in the sample. 
This categorical difference between doped and undoped topological phase transitions suggests a quantitative difference that can be explored experimentally. 

Last, we comment on the role of weak repulsive Coulomb interactions that are present in each of the TI materials previously mentioned, though we have ignored them in this study that focuses on the effects of disorder.
First, if the metal-insulator transitions remain continuous in the presence of interactions, there are several interesting effects worth considering. The first is to incorporate charged disorder~\cite{adam2007self,Skinner2014} to go beyond the short-ranged disorder we have considered so far. This will effectively dope the Dirac cones and we expect this will lead to a broadened, even larger intervening metallic phase. Second, at the same time if the interactions are screened to make them sufficiently local, the non-perturbative quasi-localized resonances will pay a large interaction energy cost and therefore we expect local interactions to suppress these rare regions and should narrow the metallic sliver. It will be fascinating to study these competing effects in future work.
 
\label{sec:conclusion}

\acknowledgements{We thank Sankar Das Sarma for useful discussions and collaborations on related work. Y.F.\ and J.H.P.\ are partially supported by NSF CAREER Grant No. DMR1941569 and the Alfred P. Sloan Foundation through a Sloan Research Fellowship.  J.H.W.\ acknowledges support from NSF CAREER grant DMR-2238895. D.A.H.\ was supported in part by NSF QLCI grant OMA-2120757.  Part of this work was performed at the Aspen Center for Physics, which is supported by National Science Foundation grant PHY-2210452 (J.H.W, J.H.P.) as well as the Kavli Institute of Theoretical Physics that is supported in part by the National Science Foundation under Grants No.~NSF PHY-1748958 and PHY-2309135 (J.H.W, J.H.P.).
The authors acknowledge the following research computing resources that have contributed to the results reported here: 
the Open Science Grid~\cite{osg07,osg09}, which is supported by the National Science Foundation award 1148698, and the U.S.\ Department of Energy's Office of Science,
the Beowulf cluster at the Department of Physics and Astronomy of Rutgers University; and the Office of Advanced Research Computing (OARC) at Rutgers, The State University of New Jersey (http://oarc.rutgers.edu), for providing access to the Amarel cluster.
The Flatiron Institute is a division of the Simons Foundation.}

\appendix

\section{Pushing states away from zero energy}
\label{sec:A1}
In this Appendix, we discuss some useful details that are important to accurately get the rare region contribution to the conductivity. This is based on the approach described in Ref.~\cite{Pixley-2016}.

First, consider the leading perturbative correction to the energy eigenvalues in the disorder potential. 
This is equal to $\sum_{\br}V(\br)\sim W/L^{3/2}(\mathrm{random \,\, sign})$, and even though it averages to zero, it broadens any features in $\rho(E)$ or $\sigma(E)$ to the leading order.
We remove this perturbative correction by ensuring that each sample has a random potential that sums exactly to zero. 
This amounts to working with the shifted potential $\tilde{V}(\br)=V(\br)-L^{-3}\sum_{\br'}V(\br')$.

\begin{figure}[ht]
\centering
\includegraphics[width=\columnwidth]{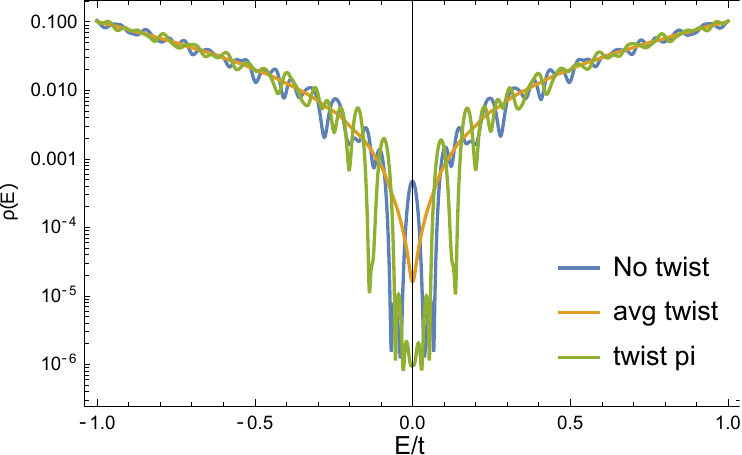}
\caption{Comparison of three choices of the treatment of the boundary condition (periodic boundary conditions, averaging over twisted boundary conditions, and a twist of $(\pi,\pi,\pi)$   
labeled ``no twist'', ``avg twist'', ``twist pi,'' respectively)
on the finite size density of states in the clean limit (i.e., no disorder) at the Dirac semimetal critical point $m_2/m_0=-2$. Shown is a KPM expansion order of $N_C=2^{10}$ and a linear system size of $L=60$.}
\label{fig:rho_bcs}
\end{figure}

We now focus on introducing the largest finite size gap possible in the numerics to allow for rare region effects to dominate near low energy.
To do so we apply twisted boundary conditions with a twist of $\bm{\theta}=(\pi,\pi,\pi)$ to push the low energy states as far away zero energy as possible.
As can be seen in the density of states in Fig.~\ref{fig:rho_bcs}, the twist creates a finite-size gap and averaging over random twists gives a smooth interpolation through this gap, and just applying periodic boundary conditions produces a large finite size effect due to the states at or near zero energy.
We take advantage of this appendix when computing the conductivity along the perturbative semimetal line.

\section{Finite energy level statistics}
\label{sec:A2}

\begin{figure}[ht]
\centering
\includegraphics[width=\columnwidth]{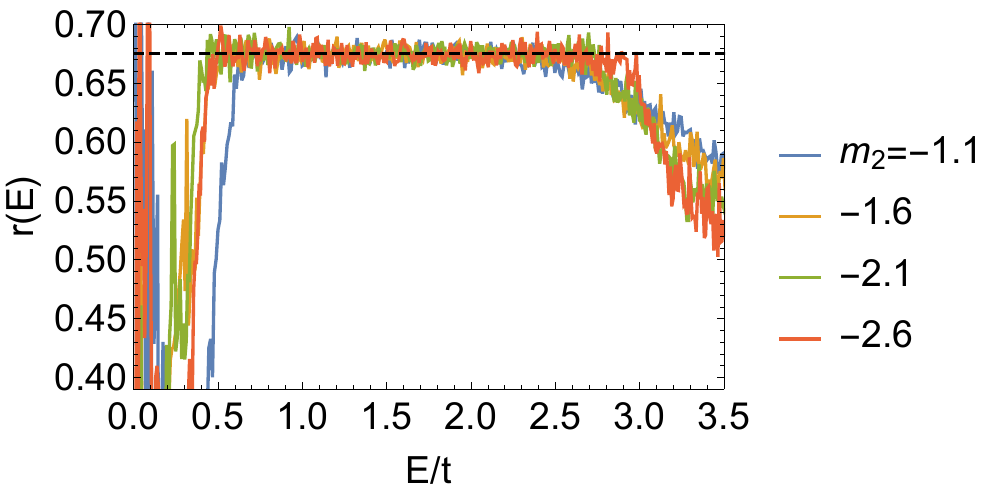}
\includegraphics[width=\columnwidth]{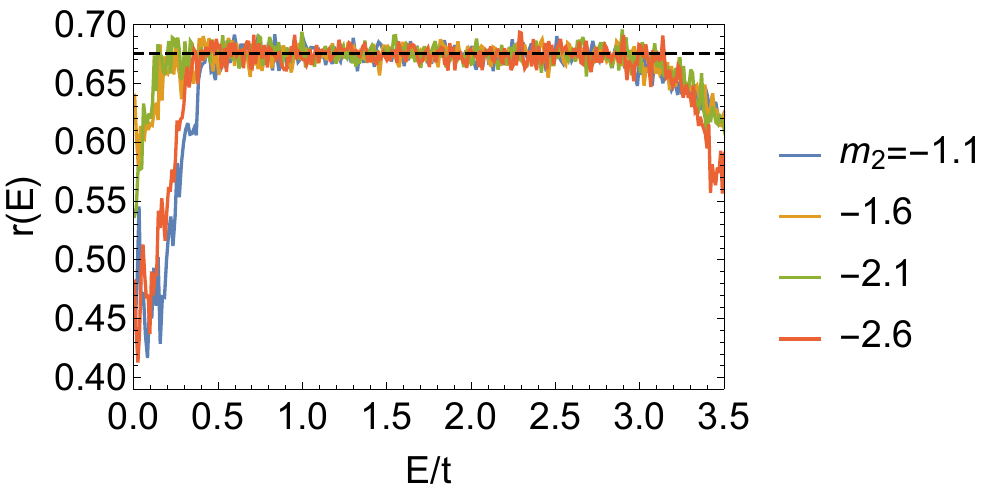}
\caption{Average adjacent gap ratio $r(E)$ as a function of energy for $W=0.7$ (top) and $W=1.0$ (bottom) for several values of $m_2$, for $L=10$ computed using exact diagonalization. The data shows that the finite energy level statistics clearly follows the GSE prediction of $r_\mathrm{GSE}\approx0.6750$~\cite{Atas-gapratio-2013} shown as a black dashed line.}
\label{fig:level_stats}
\end{figure}
In this appendix we study the level statistics of the diffusive metal phase. Here, we present results on the energy resolved adjacent gap ratio 
\begin{equation}
    r(E_i) = \left[ \frac{
    \mathrm{max}(\delta_{i},\delta_{{i+1}})
    }
    {\mathrm{min}(\delta_{i},\delta_{{i+1}})}
    \right ]
\end{equation}
where $\delta_i = E_i-E_{i-1}$,
that is a dimensionless measure of the 
level statistics of the model. We compute this from the full spectrum and therefore focus  on small sizes using exact diagonalization, with periodic boundary conditions. For the data presented here, we averaged over 200 samples. In Fig.~\ref{fig:level_stats}, we fix the disorder strength and vary $m_2$ for $W=0.7$ and $W=1.0$ that are below and above the avoided transition. Due to the small finite size, we find it challenging to accurately resolve the low-energy level statistics near the band center. At finite energy however, we find very nice agreement with the Gaussian Symplectic Ensemble of random matrix theory.

\bibliography{refs2}

\end{document}